%% file: no_comments.tex
\documentclass[screen,sigconf]{acmart}

 \setcopyright{none}
 \setcopyright{none}
 \renewcommand\footnotetextcopyrightpermission[1]{}

\usepackage{enumitem}
\usepackage{flushend}
\usepackage{booktabs}
\usepackage{fixmath}
\usepackage{pifont}
\usepackage{multirow}
\usepackage{graphicx}  %
	\setkeys{Gin}{width=\textwidth, height=\textheight, keepaspectratio}
\usepackage{subcaption}
\newif\ifcomment
\commentfalse
\ifcomment
	\newcommand{\edc}[1]{\textbf{\em\color{red}EDC: #1}}
	\newcommand{\gvg}[1]{\textbf{\em\color{blue}GG: #1}}
\else
	\newcommand\edc[1]{}
	\newcommand\gvg[1]{}
\fi

\newcommand{\descr}[1]{\smallskip\noindent\textbf{#1}}

\renewcommand{\footnoterule}{%
  \kern -3pt                 %
  \hrule width 0.4\columnwidth %
  \kern 2.6pt                %
}
\usepackage[hang,flushmargin]{footmisc}

\begin{document}
\pagestyle{plain}

\sloppy

\title{Rethinking Anonymity Claims in Synthetic Data Generation:\\A Model-Centric Privacy Attack Perspective$^*$}

\author{Georgi Ganev$^{1}$, Emiliano De Cristofaro$^{2}$\\[0.5ex]
\large $^1$University College London and SAS, $^2$University of California, Riverside\\[3ex]
}

\thanks{$^*$Published in the Proceedings of the 25th Workshop on Privacy in the Electronic Society, WPES 2026, part of ACM CCS 2026.}

\renewcommand{\shortauthors}{Georgi Ganev and Emiliano De Cristofaro}
\renewcommand{\shorttitle}{Rethinking Anonymity Claims in Synthetic Data Generation}

\begin{abstract}
Training generative machine learning models to produce synthetic tabular data has become a popular approach for enhancing privacy in data sharing.
As this typically involves processing sensitive personal information, releasing either the trained model or generated synthetic datasets can still pose privacy risks.
Yet, recent research, commercial deployments, and privacy regulations like the General Data Protection Regulation (GDPR) largely assess anonymity at the level of an individual dataset.

In this paper, we rethink anonymity claims about synthetic data from a model-centric perspective, arguing that meaningful assessments must account for the underlying generative model and be grounded in state-of-the-art privacy attacks.
This perspective better reflects real-world deployments, where trained models are often accessible for interaction or querying.
We interpret the GDPR's definitions of personal data and anonymization under such access assumptions to identify the identifiability risks that must be mitigated and map them to privacy attacks across threat settings.
We then argue that synthetic data techniques alone do not ensure sufficient anonymization.
Finally, we compare the two mechanisms most commonly used with synthetic data -- Differential Privacy (DP) and Similarity-based Privacy Metrics (SBPMs) -- and argue that while DP can offer robust protections against identifiability risks, SBPMs lack adequate safeguards.
Overall, our work connects regulatory notions of identifiability with model-centric privacy attacks, enabling more responsible and trustworthy assessment of synthetic data systems by researchers, practitioners, and policymakers.
\end{abstract}

\maketitle

\section{Introduction}
\label{sec:intro}
The concept of {\em synthetic data} is increasingly being advocated as a privacy-enhancing approach to sharing sensitive tabular data, e.g., by the UN~\cite{un2023guide}, the Royal Society~\cite{rs2023privacy}, and the OECD~\cite{oecd2023emerging}.
Governmental initiatives in the US~\cite{abowd2022the}, UK~\cite{ons2023synthesising}, and Israel~\cite{hod2025differentially} have begun using synthetic data generation to release census data, while the European Commission (EC) has been conducting feasibility studies in the context of synthetic health and financial data~\cite{girolamo2024synthetic,hradec2022multipurpose}.
Commercial offerings from the private sector are also growing rapidly, both in terms of investments~\cite{forbes2022synthetic,techcrunch2022the} and products in financial, insurance, and healthcare settings, where models are trained on large sensitive datasets~\cite{gretel2022gretel,mostly2022leading,tonic2022getting}.
As a consequence, regulators have started to consider synthetic data as a privacy-enhancing tool for sharing sensitive data~\cite{fca2024using,ico2022privacy,ico2023privacy} though falling short of issuing specific regulatory guidance.

The intuition behind synthetic data is to train generative machine learning models\footnote{While there are other methods for generating synthetic data, like statistical and rule-based approaches and agent-based simulations, we focus on generative machine learning models since they are the most popular and successful approach~\cite{cristofaro2024synthetic,hu2024sok,jordon2022synthetic}.} -- e.g., Bayesian networks~\cite{zhang2017privbayes}, Generative Adversarial Networks~\cite{jordon2018pate,xu2019modeling}, Diffusion Models~\cite{kotelnikov2023tabddpm,zhang2024mixed}, Transformer-based~\cite{borisov2023language,solatorio2023realtabformer} -- to capture the overall patterns and correlations in a sensitive dataset and generate new records that preserve individual privacy while reflecting the same aggregate patterns.
Modern synthetic data deployments resemble other generative AI products, emphasizing interactive access to the underlying model and promoting the ability to create unlimited synthetic datasets as well as conditional generation for bias correction and dataset rebalancing~\cite{gretel2024gretel,mostly2024truly}.

\descr{Motivation.}
Extensive research has demonstrated that generative models can be vulnerable to privacy attacks, as they may memorize and reproduce individual records in the synthetic data~\cite{carlini2023extracting,carlini2019secret,carlini2021extracting,nasr2023scalable,webster2019detecting}.
These vulnerabilities raise serious concerns, particularly in sensitive domains like healthcare and finance, and risk undermining public trust in synthetic data as a privacy-enhancing technology.
Nonetheless, privacy attacks on generative models have been mainly overlooked by machine learning researchers and commercial deployments, which often rely on ad-hoc, dataset-specific evaluations to justify privacy and anonymity claims.
For instance, recently proposed Diffusion/LLM-based models, published in top-tier machine learning venues~\cite{kotelnikov2023tabddpm,mueller2025continuous,pang2024clavaddpm,shi2025tabdiff,zhang2024mixed,wang2024harmonic}, exclusively use privacy metrics based on similarity to the original data (which we refer to as Similarity-based Privacy Metrics~{\em (SBPMs)}~\cite{ganev2025inadequacy,platzer2021holdout}) to argue that the generated synthetic datasets are private.

While some commercial solutions offer training models that satisfy rigorous privacy definitions, such as Differential Privacy~{\em (DP)}~\cite{dwork2006calibrating,dwork2014algorithmic}, the majority rely on the same ad-hoc SBPMs~\cite{ganev2025inadequacy}.
Regardless of the privacy mechanism used, commercial solutions assert compliance with data protection laws, most commonly referencing anonymization standards under the General Data Protection Regulation (GDPR)~\cite{official2016recital}.
These regulatory claims, however, remain largely unverified.
More fundamentally, current data protection frameworks like the GDPR remain largely dataset-centric, overlooking the {\em model-centric} nature of modern synthetic data products.

\descr{Roadmap.}
In this paper, we aim to bridge the gap between the privacy protections offered by synthetic data techniques and relevant regulations.
We focus on the GDPR in the EU and the UK, and more specifically, on data anonymization requirements.
First, we interpret the legal definitions of personal data and the standards for anonymization from a machine learning and model-centric perspective, as these were originally framed with a more traditional, database-centric view.
Then, we examine current synthetic data solutions, privacy attacks aimed at extracting personal information from trained models and synthetic data, and technologies designed to minimize these risks, discussing these issues in the context of legal requirements.

\descr{Contributions.}
In summary, we argue that claims about synthetic data anonymity should account for the capabilities of the underlying generative model and be evaluated using state-of-the-art privacy attacks.
Our contributions include:
\begin{enumerate}
  \item We interpret the three key risks that must be minimized for sufficient anonymization under the GDPR -- namely, singling out, linkability, and inferences -- in the context of generative models and synthetic data generation. %
  \item We operationalize the three risks through known privacy attacks, where a successful attack provides evidence that the corresponding risk has not been sufficiently mitigated.
	More precisely, we map singling out to differencing attacks, linkability to membership inference, and inferences to attribute inference.
	We frame our analysis using the ``motivated intruder'' test, applying varying levels of capability and strictness to the attacks. %
	\item We argue that synthetic data techniques, on their own, do not mitigate the regulatory risks.
	We then compare and contrast the privacy protections offered by DP and SBPMs,
	arguing that DP has the potential to mitigate all risks and achieve sufficient anonymization in appropriate contexts, while SBPMs do not provide adequate protections and are unlikely to meet regulatory standards. %
\end{enumerate}

Overall, our work connects current data protection regulations (i.e., GDPR) with the scientific literature on synthetic data generation and privacy attacks.
By explicitly accounting for the generative model's role in synthetic data systems, our analysis provides researchers, practitioners, and policymakers with a model-centric perspective to assess privacy and regulatory risks in ways that better reflect real-world deployment.
Such an approach can promote more accountable and trustworthy data sharing while helping users better understand and trust the processes underlying synthetic data generation.
Finally, our work lays the groundwork for future research examining other aspects of the GDPR, such as data collection (purpose limitation), data processing consent, data minimization, data access and governance, supporting more comprehensive, regulation-aligned synthetic data practices.

\section{Preliminaries}
\label{sec:prelim}
This section reviews background notions related to synthetic data, generative machine learning models, synthetic data deployments, and regulatory claims, as well as relevant privacy attacks.

\subsection{Synthetic Data and Generative Models}
\label{subsec:synthetic}
The term ``synthetic data'' typically refers to artificial data, not produced by real-world events.
In this paper, we focus specifically on synthetic tabular data and on generative machine learning models, which are the most widely used approach for producing such data in both research and practice~\cite{cristofaro2024synthetic,hu2024sok,jordon2022synthetic}.
The term is also often used informally to describe the techniques used to generate synthetic data.
Overall, synthetic data can be considered a subfield of generative AI, encompassing algorithms that are generally trained in a centralized manner, though on a much smaller scale than modern Large Language Models~(LLMs).

\begin{figure}[t!]
	\centering
	\includegraphics[width=0.6\linewidth]{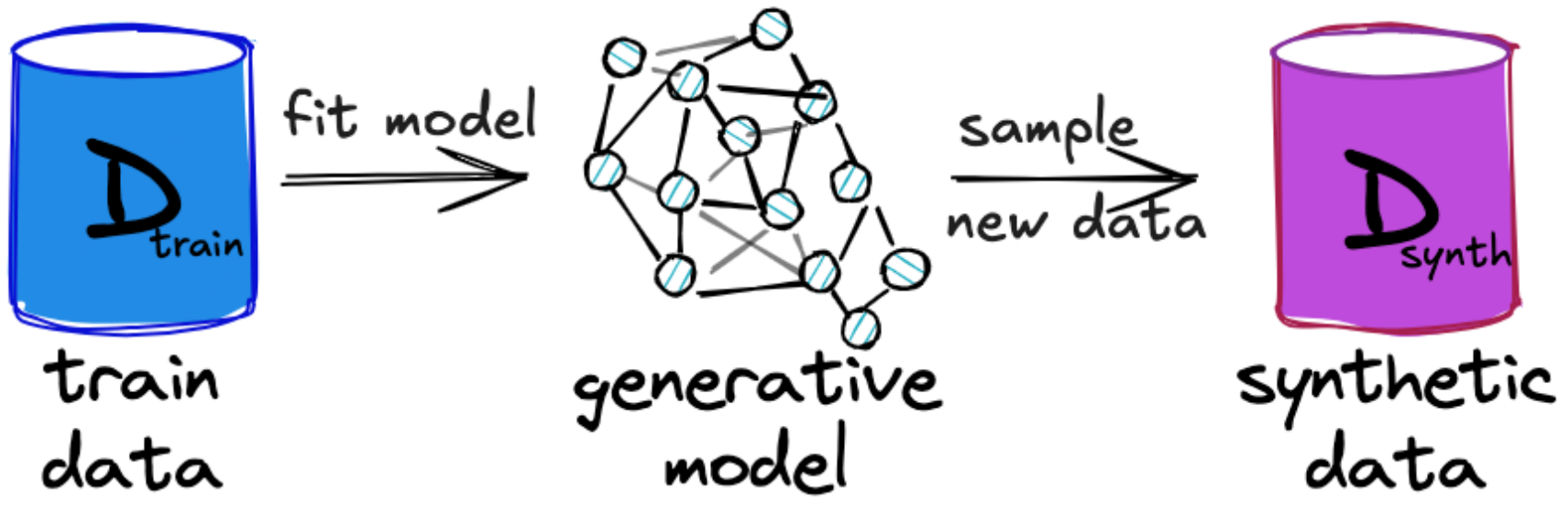}
	\caption{Synthetic data generation using generative models.}
	\label{fig:syn}
\end{figure}

\descr{Definition.}
As illustrated in Figure~\ref{fig:syn}, we consider a generative model $\mathcal{G}$ taking in input $\mathcal{D}_{train}$, a dataset of $n$ individuals' records.
During training, $\mathcal{G}$'s parameters are updated to learn a (lower-dimensional) representation of the underlying data probability distribution.
The model can then be (repeatedly) sampled to generate new, synthetic datasets.

\descr{Synthetic Data Techniques.}
Numerous approaches have been proposed, ranging from statistical methods to deep learning techniques.
The former include copulas~\cite{gambs2021growing,li2014differentially}, graphical models~\cite{cai2021data,mahiou2022dpart,mckenna2021winning,zhang2017privbayes}, and workload/query-based ones~\cite{aydore2021differentially,liu2021iterative,mckenna2022aim,vero2024cuts,vietri2022private,vietri2020new,zhang2021privsyn}.
The latter typically include Variational Autoencoders (VAEs)~\cite{abay2018privacy,acs2018differentially,takagi2021p3gm}, Generative Adversarial Networks (GANs)~\cite{frigerio2019differentially,jordon2018pate,long2021gpate,tantipongpipat2021differentially,xie2018differentially,xu2019modeling,zhang2018differentially}, Diffusion Models~\cite{kotelnikov2023tabddpm,mueller2025continuous,pang2024clavaddpm,shi2025tabdiff,zhang2024mixed}, and Transformer/LLM-based~\cite{borisov2023language,castellon2024dp,solatorio2023realtabformer,wang2024harmonic}.\footnote{For more in-depth surveys and benchmarks on synthetic data generation, please refer to~\cite{cristofaro2024synthetic,du2024towards,ganev2024graphical,hu2024sok,jordon2022synthetic,tao2022benchmarking,chen2025benchmarking,qi2026clover}.}

\subsection{Synthetic Data Deployments} %
\label{sec:comp}
The synthetic data market has been experiencing steady growth in recent years.
As with other emerging technologies, startup companies are often well-positioned in the early stages to drive innovation and take on significant risks; therefore, the landscape has been dominated mainly by startups, several of which have been acquired by larger firms~\cite{aetion2022aetion,nasdaq2022anonos,sas2024sas,techcrunch2025nvidia}.

Overall, a review of the synthetic data ecosystem reveals that there currently are over ten well-funded providers~\cite{forbes2022synthetic,techcrunch2022the} that deliver production-ready synthetic data solutions to businesses across various industries.
Notable applications include large enterprises leveraging these providers' technologies to facilitate data sharing both internally and externally in e-commerce~\cite{tonic2022getting}, insurance~\cite{mostly2022leading}, and healthcare~\cite{gretel2022gretel}. %
Moreover, regulators in the UK, like the ICO and FCA, have highlighted some of these deployments in the financial sector~\cite{fca2024using,ico2023synthetic}.

\descr{Workflow.}
The workflow of early synthetic data technologies (between 2018 and 2022) largely followed the principles of traditional anonymization techniques, like k-anonymity~\cite{sweeney2002k}, l-diversity~\cite{machanavajjhala2007diversity}, t-closeness~\cite{li2006t}.
In other words, the generative model served as an intermediate step to produce a single synthetic dataset that replaced the sensitive data, after which the model itself was not exposed to end users and discarded, keeping the primary focus on the dataset.

However, the rise of generative AI products has significantly influenced the workflow of synthetic data technologies.
Modern deployments increasingly prioritize the generative model itself, allowing users to retrain/adapt models and interact with them directly via API calls or user interfaces~\cite{gretel2024gretel,mostly2024truly}.
Even when large pre-trained models like LLMs are used as the basis for generation, rather than training a model from scratch, domain adaptation/fine-tuning on private/sensitive data is still often necessary to obtain high utility, because public data may not be representative of the sensitive data distribution~\cite{tramer2024position}.
This shift enables the generation of multiple or even unlimited synthetic datasets, along with advanced capabilities like conditional data generation to correct biases, rebalance datasets, and enhance signal strength.

\descr{Compliance Claims.}
Although no established standards exist for synthetic data compliance, providers often claim their solutions meet anonymization requirements under regulations like the GDPR, as documented in recent work~\cite{ganev2025inadequacy}.

\subsection{Privacy Attacks against Synthetic Data}
\label{subsec:attacks}
While releasing synthetic data appears inherently ``{\em privacy friendlier}'' than disclosing real data, it does not guarantee that an adversary cannot infer sensitive information about the training data.
For example, generative models may overfit to %
or memorize specific training records, potentially reproducing them either exactly or approximately during data generation~\cite{van2021memorization,carlini2023extracting,carlini2019secret,carlini2021extracting,nasr2023scalable,webster2019detecting}.
In fact, sometimes memorization -- especially of outliers %
-- may be necessary when high utility is a strict  requirement~\cite{feldman2020does}.
Next, we introduce well-known privacy attacks. %

\begin{figure}[t!]
	\centering
	\includegraphics[width=0.6\linewidth]{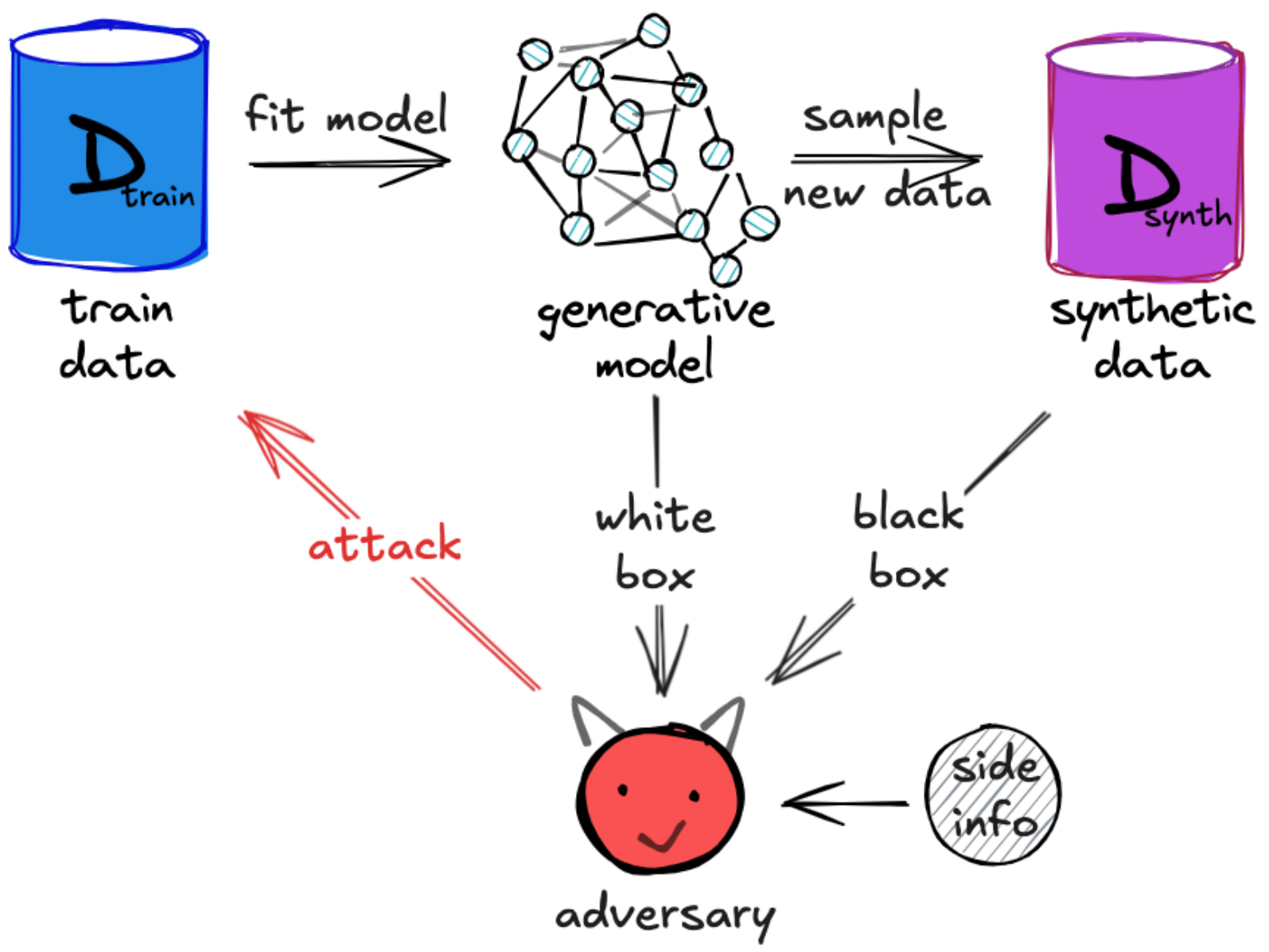}
	\caption{Privacy attacks on synthetic data: an adversary with black or white-box access to the model (and side information) extracts private information from the training data.}
	\label{fig:attack}
\end{figure}

\descr{Adversarial Model.}
We assume a strategic adversary aiming to extract sensitive information about the training data.
Their capabilities and assumptions may vary, leading to different threat models.
For instance, an adversary might have limited access, such as only having API access to a trained generative model (aka {\em black-box} settings), or they might have full access to the model, including its parameters and architecture (aka {\em white-box} settings).
The adversary may also possess additional side information, e.g., knowledge of the underlying algorithm, access to similar datasets, or prior knowledge about the data distribution.
In many attack scenarios, the adversary focuses on a target record, which is usually a specific data point from the training dataset they aim to identify, infer attributes about, or reconstruct entirely.
We illustrate the general settings of black- and white-box attacks in the context of synthetic data in Figure~\ref{fig:attack}.

\descr{Common Attacks.}
In the context of synthetic data generation, Membership Inference Attacks (MIAs)~\cite{van2023membership,chen2020gan,guepin2023synthetic,hayes2019logan,hilprecht2019monte,houssiau2022tapas,stadler2022synthetic} involve an adversary that attempts to infer whether or not a given record was used to train the generative model.
MIAs represent an intuitive type of attack and are commonly used as a measuring stick to assess whether the generative models are ``leaky.''
Another standard set of attacks is Attribute Inference Attacks (AIAs)~\cite{annamalai2024linear,houssiau2022tapas,stadler2022synthetic}, where an adversary with partial knowledge of some training records seeks to infer sensitive, unknown feature(s).
Finally, Reconstruction attacks~\cite{carlini2023extracting,carlini2021extracting,ganev2025inadequacy,nasr2023scalable} aim to fully recover or extract entire train records by using the trained model.\footnote{For more in-depth surveys on privacy attacks, please refer to~\cite{de2021critical,dwork2017exposed,jegorova2022survey,papernot2018sok,rigaki2023survey}.}

\section{Synthetic Data Privacy Mechanisms}
\label{sec:mech}
In this section, we review the two privacy mechanisms most commonly used in the context of synthetic data generation -- Differential Privacy~(DP) and Similarity-based Privacy Metrics~(SBPMs).
While SBPMs are not a true privacy mechanism, as they have no formal privacy definition before accessing the data, we nevertheless refer to them as such because they are commonly used with the intent of providing privacy.

\subsection{Differential Privacy (DP)}
\label{subsec:DP}
Differential Privacy~(DP) is a mathematical definition that limits the information that can be inferred about {\em any} specific individual from analyzing or computing on their data.
Put simply, DP bounds the probability of determining whether any individual's data was included in the input of an algorithm by examining its output.
In the context of machine learning, DP limits the success of an adversary with access to a trained model in inferring whether or not the data of a person was used to train that model.

\descr{Definition.}
A randomized algorithm $\mathcal{A}$ satisfies ($\epsilon, \delta$)-DP if, for all sets of its possible outputs $S$ and all neighboring datasets $\mathcal{D}$ and $\mathcal{D}^{\prime}$, it holds~\cite{dwork2006calibrating,dwork2014algorithmic}:
\begin{equation}
	P[{\mathcal{A}}(D)\in S]\leq \exp \left(\epsilon \right)\cdot P[{\mathcal{A}}(\mathcal{D}^{\prime})\in S] + \delta
\end{equation}
The definition of neighboring datasets typically relates to two datasets that are identical except for a single individual's data, although this could vary according to the specific DP definition~\cite{desfontaines2020sok}.
The value $\epsilon$ (aka the privacy budget) is a positive real number quantifying the information leakage -- more precisely, the level of indistinguishability.
The smaller the $\epsilon$, the more private the computation.
The parameter $\delta$ is an asymptotically small real number accounting for a negligible probability of failure. %

DP also has two key properties: composition~\cite{kairouz2015composition} and post-processing~\cite{dwork2014algorithmic}.
The former allows multiple DP mechanisms to be combined while seamlessly tracking the overall privacy budget; the latter ensures that once a model is trained under DP, it can be reused any number of times without additional privacy loss.
Thus, repeated/conditional sampling and downstream training on model outputs require no additional budget, whereas retraining/fine-tuning on sensitive data incurs additional privacy loss.

\begin{figure}[t!]
	\centering
	\includegraphics[width=0.6\linewidth]{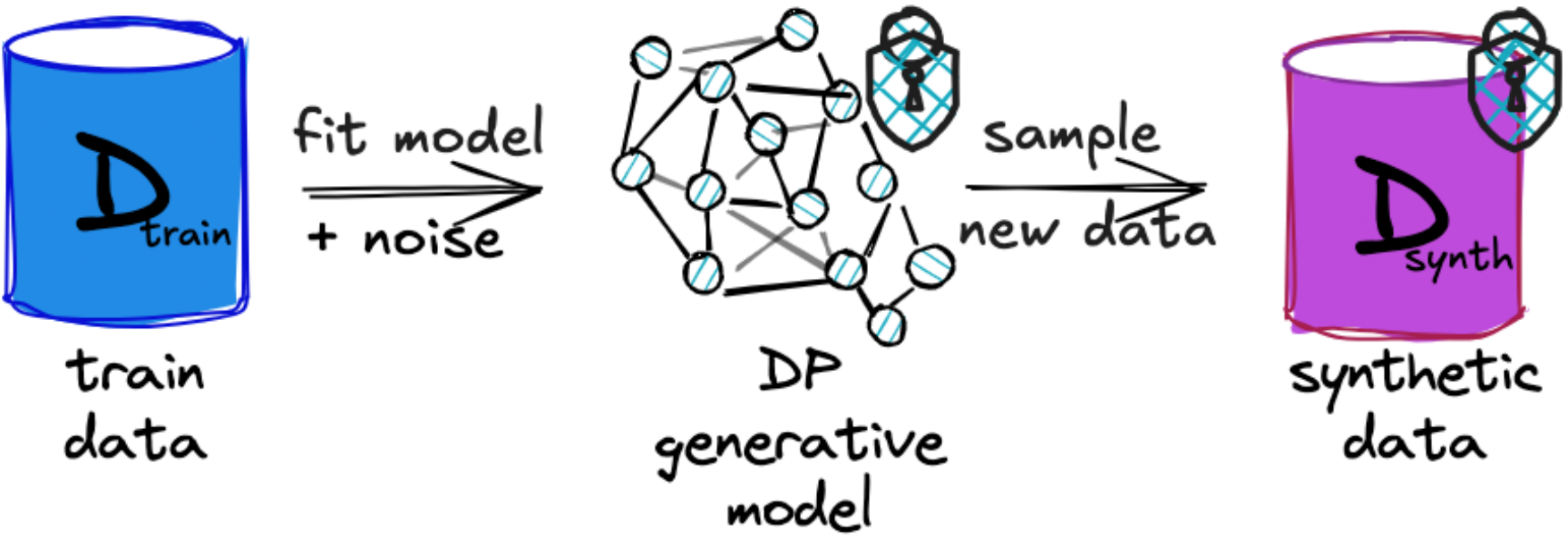}
	\caption{Synthetic data generation with Differential Privacy~(DP); model is trained while satisfying DP.}
	\label{fig:dp_syn}
\end{figure}

\descr{DP and Synthetic Data.}
In the context of synthetic data, %
DP is usually satisfied by introducing noise or randomness during model training (see Figure~\ref{fig:dp_syn}) through Laplace~\cite{dwork2006calibrating}, Gaussian~\cite{dwork2006our}, or Exponential~\cite{dwork2006calibrating} mechanisms, and using frameworks like DP-SGD~\cite{abadi2016deep} or PATE~\cite{papernot2017semi,papernot2018scalable}.
There is a vast body of literature proposing DP generative models, e.g.,~\cite{aydore2021differentially,jordon2018pate,liu2021iterative,mckenna2021winning,mckenna2022aim,vietri2020new,xie2018differentially,zhang2017privbayes,zhang2021privsyn}.
Some proposals have won synthetic data competitions organized by the NIST~\cite{nist2018differential,nist2018the} and been used by, e.g., the UK and Israel governments to release census data~\cite{hod2025differentially,ons2023synthesising}.
Around half of the leading synthetic data vendors reportedly include DP generative models in their product offerings~\cite{ganev2025inadequacy}.

\subsection{Similarity-based Privacy Metrics (SBPMs)}
\label{subsec:sbpms}
Alternative privacy heuristics used in the context of synthetic data include Similarity-based Privacy Metrics~(SBPMs).
Rather than formal definitions, these rely on ad-hoc empirical evaluations; generally speaking, they measure the privacy of a synthetic dataset by calculating the distances between datasets and running pass/fail statistical tests.
The intuition is that the synthetic data should be representable and similar to the training dataset but not too close, i.e., not as close as one would expect from the holdout test dataset~($\mathcal{D}_{test}$) that serves as a reference~\cite{mobey2022help,platzer2021holdout}.

While many ad-hoc privacy metrics have been proposed and applied~\cite{boudewijn2023privacy}, we focus on the three most widely adopted SBPMs by companies and researchers~\cite{desfontaines2024empirical,ganev2025inadequacy}: Identical Match Share (IMS), Distance to Closest Records (DCR), and Nearest Neighbor Distance Ratio (NNDR).

\begin{figure}[t!]
	\centering
	\includegraphics[width=0.45\linewidth]{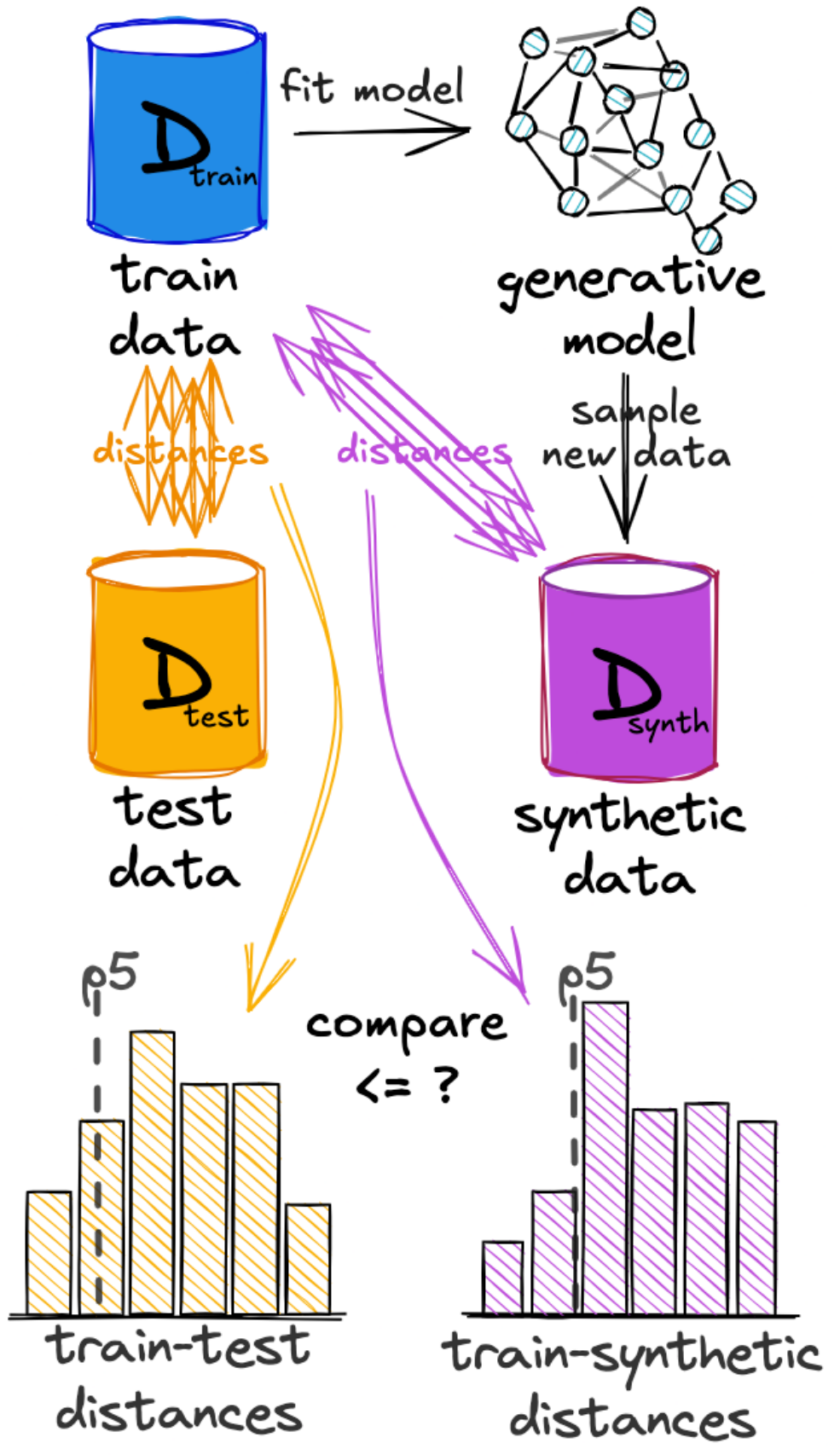}
	\caption{Synthetic data generation with Similarity-based Privacy Metrics~(SBPMs); the privacy of synthetic data is measured through SBPMs.}
	\label{fig:sbpm_syn}
\end{figure}

\descr{Definition.}
The input to all three metrics consists of $\mathcal{D}_{train}$, $\mathcal{D}_{synth}$, and $\mathcal{D}_{test}$.
Note that $\mathcal{D}_{test}$ is the same size as $\mathcal{D}_{train}$ and comes from the same distribution but is not used to train the generative model.
As illustrated in Figure~\ref{fig:sbpm_syn}, the closest pairwise distances from $\mathcal{D}_{synth}$ to $\mathcal{D}_{train}$, $d_{synth} = d(\mathcal{D}_{train}, \mathcal{D}_{synth})$ (i.e., for each synthetic record, the distance to its closest training record/neighbor is calculated) and $d_{test} = d(\mathcal{D}_{train}, \mathcal{D}_{test})$ are computed and their distributions compared through a statistical test.
The passing criterion is based on comparing a simple statistic from each distribution, such as the average or 5th percentile~(p5).
Finally, $\mathcal{D}_{synth}$ is deemed private if all three privacy tests pass~\cite{mostly2020truly,panfilo2022generating}.

\descr{Identical Match Share (IMS)}
calculates the proportion of exact copies between records in $\mathcal{D}_{train}$ and $\mathcal{D}_{synth}$ (statistic: average; test: $d_{test} \geq d_{synth}$).
IMS allows for some synthetic records to be identical to those in the training data, but only up to a specified limit to ensure privacy.

\descr{Distance to Closest Records (DCR)}
calculates the two distances $d_{test}$ and $d_{synth}$ using the nearest neighbors (statistic: p5; test: $d_{test} \leq d_{synth}$).
DCR is designed to prevent scenarios where slightly perturbed training data is passed off as synthetic.

\descr{Nearest Neighbor Distance Ratio (NNDR)}
is similar to DCR but divides the nearest neighbor distances by the distance to the second nearest neighbor (statistic: p5; test: $d_{test} \leq d_{synth}$).
This ratio further protects against privacy leakage, particularly for outliers in the training data.

\descr{SBPMs and Synthetic Data.}
Numerous generative models proposed in literature, including in domains like medicine, rely solely on SBPMs as their privacy mechanism~\cite{borisov2023language,damico2023synthetic,guillaudeux2023patient,liu2023tabular,lu2019empirical,panfilo2022generating,park2018data,solatorio2023realtabformer,yoon2023ehr,zhao2021ctab}, etc.
Recent Diffusion/LLM-based papers published in top-tier machine learning venues~\cite{kotelnikov2023tabddpm,mueller2025continuous,pang2024clavaddpm,shi2025tabdiff,zhang2024mixed,wang2024harmonic} similarly only use SBPMs -- or more precisely DCR -- to claim the generated synthetic datasets are privacy-preserving.
In fact, the EC used SBPMs to assess the privacy of synthetic data generated from cancer patient records~\cite{hradec2022multipurpose}.
Additionally, half of the leading synthetic data vendors depend exclusively on SBPMs to measure and actively demonstrate the privacy of synthetic datasets to their clients~\cite{ganev2025inadequacy}.

To privacy researchers, %
it might appear %
odd that leading synthetic data companies provide clients with unperturbed access to the privacy metrics scores for each generated synthetic data that require direct queries/computations on the sensitive training data.
In theory, one could argue that restricting access to the metrics or adding noise to them through DP mechanisms would alleviate the issue, but these are not robust solutions, as argued by~\cite{ganev2025inadequacy}.
Disabling metrics access would undermine a key selling point for providers, as these metrics serve as tangible proof of compliance, crucial for transparency and explainability; without them, users would be forced to rely on blind trust in the provider.
Applying DP to the metrics would also require allocating additional privacy budget for each generation run, clashing with one of the primary advantages of synthetic data, i.e., the ability to generate unlimited data.  %

\section{Connecting Regulation, Synthetic Data, and Privacy Attacks}
We outline the regulatory definitions and key risks associated with processing personal data, interpret them in the context of generative models and synthetic data, and map these risks to privacy attacks.

\subsection{Regulatory Definitions}
\label{subsec:reg}

\descr{Personal Data.}
Under data protection law, information is generally assessed by whether or not it meets the definition of personal data~\cite{official2016article,official2016recital}. %
In the EU and UK, personal data is defined as ``any information relating to an identified or identifiable living individual,'' whereby ``identifiable living individual'' is someone who can be identified (directly or indirectly) by reference to an identifier such as a name, ID number, or physical, genetic, social identity~\cite{official2016article}.
Information that is effectively anonymized is not considered personal data, and data protection law does not apply to it~\cite{official2016recital,ico2021how}.
In practice, however, the actual identifiability of individuals can be highly context-specific as different types of information carry different levels of identifiability risks depending on the circumstances.

Creating synthetic data via generative machine learning techniques that use sensitive personal data\footnote{As discussed in Section~\ref{sec:comp}, in practice, organizations train generative models on large sensitive datasets containing customer data, which pertains to living individuals who can be identified and thus meets the criteria for personal data.} as input, {\em does} require processing it.
However, whether the resulting trained model---and, by extension, the synthetic data---constitutes personal data or anonymous information is to be determined by assessing the identifiability risk, considering the relevant context.
This raises the question of what constitutes a sufficient level of anonymization, which we discuss next.

\descr{Sufficient Anonymization.}
The Information Commissioner's Office (ICO), the UK's main regulatory and national data protection authority, states that ``effective anonymization reduces identifiability risk to a sufficiently remote level''~\cite{ico2021how}.
When assessing whether someone is identifiable, one should take into account any ``means reasonably likely to be used.''
Objective factors to consider include the cost and time required to identify, the available underlying technologies, technological development over time, etc.
However, not every hypothetical or theoretical chance of identifiability needs to be considered.
Instead, the focus should be on what is reasonably likely to be used, given the circumstances, rather than in absolute terms.
This aligns with the approach of Article 29 Data Protection Working Party~\cite{eu2014opinion}, which also notes that data controllers should regularly reassess the attendant risks.

Unlike traditional anonymization techniques, which take a database-centric approach and assume a one-to-one dependency between rows in the input and output datasets, we argue that risk assessment in synthetic data generation should instead be {\em model-centric.}
The primary focus should be on the trained model and, by extension, the potential synthetic datasets it can generate.
This is because: 1) the model processes personal data during training,\footnote{There are also suggestions that models trained on personal data might themselves be considered personal data~\cite{veale2018algorithms}.} and 2) the extent to which the model memorizes or retains personal data is unclear, posing a risk of reproducing sensitive information in the synthetic data.
Moreover, adopting a model-centric risk assessment aligns with real-world deployments of synthetic data technologies, as discussed in Section~\ref{sec:comp}.
It has also been recommended by European Data Protection Board~\cite{eu2024opinion} to assess the anonymity of AI models.

In terms of specific assessments, Article 29~\cite{eu2014opinion} and the ICO~\cite{ico2021how} assert that the following {\em three key identifiability risks} must be mitigated to achieve sufficient anonymization: %
\begin{enumerate}[leftmargin=0.5cm]
	\item {\em Singling Out/Individuation:} an individual being isolated or distinguished from others.
	We interpret singling out as the ability to identify or isolate an individual from the training data by examining the trained model (or synthetic dataset(s)).
	For example, this could occur if the model (consistently) generates a record that closely matches a rare combination of attributes (e.g., a person's age, zip code, job title, etc.) present in the training data. %
	\item {\em Linkability:} any records/datasets (publicly available or not) being combined, thereby enabling the identification of an individual.
	We interpret linkability as the potential for the trained model (or synthetic dataset(s)) to be linked back to the training data through additional information (e.g., partial records, public/reference data, model architecture).
	For example, a synthetic dataset may exhibit patterns that, when combined with external information (e.g., demographic statistics or data from a similar distribution), could re-identify individuals. %
	\item {\em Inferences:} an attribute of an individual being deduced with significant probability from other attribute values.
	We interpret inferences as using the trained model (or synthetic dataset(s)) to infer an attribute of a training record.
	For example, whether a synthetic dataset allows accurate prediction of an individual's income based on their occupation and location, if it was part of the training data. %
\end{enumerate}

The ICO explains that the three risks should be assessed through the {\em motivated intruder} test, i.e., a competent and motivated intruder with access to appropriate resources attempting to identify individuals~\cite{ico2021how}.
This test is highly relevant in the context of synthetic data generation, where an intruder might use publicly available data/tools, model outputs, or other side information to re-identify individuals from seemingly anonymized synthetic data.
Additionally, the ICO's approach to anonymization is technology-agnostic, meaning that all privacy risks and assessments apply universally, regardless of the specific synthetic data generation technique used.

\descr{Framing Risk as Attribute Disclosure.}
One should exercise caution when considering inferences in the context of synthetic data.
On the one hand, preventing {\em statistical} inferences would contradict the goal of high-quality synthetic data generation, which aims to preserve the statistical distributions of real data.
In fact, a few privacy researchers have argued against treating statistical inferences as privacy violations~\cite{bun2021statistical,mcsherry2016statistical}.
On the other hand, if the synthetic data reveals significantly more information about an individual and their attributes when they are included in the input dataset compared to when they are not, this indicates a privacy violation~\cite{jordon2022synthetic,yeom2018privacy}.
Thus, in the rest of this paper, we discuss the inference risk from the perspective of individuals' attribute disclosure.

\subsection{From Regulatory Risks to Privacy Attacks}
\label{subsec:emp}
We now operationalize the three key regulatory risks -- singling out, linkability, and inferences -- through established empirical privacy attacks.
These mappings are non-exhaustive, but a successful attack provides evidence of the corresponding regulatory risk.
We adopt this model-level framing because it provides concrete, testable state-of-the-art attacks from the privacy literature that approximate how a motivated intruder could exploit generative models and their synthetic data outputs in practice.
However, our framework does not prescribe universal attack-success thresholds for legal anonymization, which remain context-dependent.

While privacy metrics exist for quantifying privacy protections~\cite{boudewijn2023privacy,wagner2018technical}, we argue that they inadequately address regulatory concerns.
More precisely, they lack a direct connection to specific regulatory risks and/or fail the motivated intruder test, as they only assess specific samples of training/synthetic dataset and as a result greatly underestimate the model's privacy leakage~\cite{desfontaines2024empirical,ganev2025inadequacy,yao2025dcr}.

By contrast, adversarial attacks within well-defined threat models~\cite{de2021critical,rigaki2023survey} offer a more rigorous and reliable framework for assessing privacy leakage.
These attacks not only closely correspond with the motivated intruder test but also provide a more systematic means to evaluate privacy risks under diverse assumptions and threat models~\cite{cummings2024attaxonomy}.
Furthermore, they typically target the generative model itself rather than a specific dataset, better reflecting modern synthetic data deployments and aligning with the European Data Protection Board's recent opinion on assessing anonymity in AI models~\cite{eu2024opinion}.%
\footnote{The opinion focuses on generative AI, stating that anonymity assessments should ``consider direct access to the model'' and incorporate state-of-the-art privacy attacks, e.g., attribute or membership inference, and reconstruction attacks~\cite{eu2024opinion}. However, it does not explain how these attacks relate to the three regulatory risks, what kind of access to the model or side information should be assumed, or specific attacks considered. By contrast, we interpret the regulatory risks in the context of generative models/synthetic data and identify privacy attacks that capture particular manifestations of these risks, while explicitly stating all assumptions when analyzing specific attacks from the literature.}
We present these operational relationships in Table~\ref{tab:reg} and discuss specific attacks and their assumptions in our analysis in Section~\ref{sec:anon} and~\ref{sec:anondp}.
Accordingly, since attack capabilities and available technologies evolve over time, such assessments should be revisited periodically rather than treated as one-time determinations of anonymity.

\begin{table*}[t!]
	\centering
	\begin{tabular}{lcl}
		\toprule
		\bf Regulatory Risk  					& \bf Privacy Attacks Type													& \bf Specific Privacy Attack (Assumptions)																						\\
		\midrule
		Singling out   								& Differencing Attacks   														& -																																										\\
		\midrule
		Linkability  								 	& Membership Inference Attacks   										& GroundHog~\cite{stadler2022synthetic} (black-box; training algo, representative data) 				\\
		  														&    																								& Querybased~\cite{houssiau2022tapas}	(black-box; training algo, representative data)					\\
																	&    																								& DOMIAS~\cite{van2023membership}	(no-box; reference test data)												\\
																	&    																								& AuditSynth~\cite{annamalai2024what}	(white-box; training algorithm)									\\
																	&    																								& MIDST~\cite{wu2025winning}	(white/black-box; training algo, representative data)						\\
		\midrule
		Inferences   									& Attribute Inference Attacks   										& ReconSynth~\cite{annamalai2024linear} (no-box; partial training data)									\\
		\midrule
		Overall   										& Reconstruction Attacks   						              & RAP-Rank~\cite{dick2023confidence} (black-box; accurate aggregate statistics)				\\
		   														& 												 													& ReconSyn~\cite{ganev2025inadequacy}	(black-box; privacy metrics)										\\
		\bottomrule
	\end{tabular}
  \caption{Mapping between regulatory risks and privacy attack classes.
  The attacks operationalize particular manifestations of the corresponding regulatory risks, and the mappings are non-exhaustive.}
	\label{tab:reg}
\end{table*}

\descr{Singling Out $\rightarrow$ Differencing Attacks.}
To detect one manifestation of the risk of singling out, we use ``differencing'' attacks, which test whether a trained generative model exhibits individual-level sensitivity.
The adversary can probe the trained model repeatedly, with or without generic/random prompts~\cite{carlini2021extracting}, to identify outputs that appear unusually specific or persistently reproduced.
This behavior can signal that particular individuals have been isolated/memorized during the model training.
Once potential targets are identified, the adversary can refine the attack by querying the model (or relevant privacy metrics) with and without a target record and observing whether the model's behavior changes significantly.
However, mitigating such attacks does not fully eliminate singling-out risk, as other mechanisms, such as unique attribute combinations in synthetic records, may still enable individuation. %

\descr{Linkability $\rightarrow$ Membership Inference Attacks (MIAs).}
We use MIAs to test a model-level manifestation of linkability risk.
MIAs aim to infer whether a target record influenced the model's training significantly.
These attacks typically assume at least black-box access; the adversary trains shadow models on datasets that include or exclude the target record and then compares their behavior against the deployed model.
While MIAs do not capture all forms of real-world linkage (e.g., directly linking synthetic records to external databases), they provide a practical operational analogue at the model level: successful membership inference establishes a link between a target record and the model's training data.
We adopt MIAs as a non-exhaustive proxy for linkability risk, in line with previous work targeting aggregate statistics~\cite{pyrgelis2018knock}, machine learning~\cite{kumar2020mlprivacy,yeom2018privacy}, and specifically synthetic data~\cite{stadler2022synthetic}.

\descr{Inferences $\rightarrow$ Attribute Inference Attacks (AIAs).}
Arguably, AIAs are a natural fit for estimating the risk of inferring unknown attributes from known ones.
They evaluate whether a generative model leaks information about a target beyond what the adversary already knows.
As mentioned in Section~\ref{subsec:reg}, we focus on detecting excess individual information leakage when the target is part of the training data; examples of attacks adopting this approach include~\cite{annamalai2024linear,stadler2022synthetic}.
A successful AIA therefore provides evidence of inference risk through individual-level attribute disclosure.

\descr{Overall $\rightarrow$ Reconstruction Attacks.}
Finally, we also examine (full-record) reconstruction attacks, one of the most powerful classes of privacy attacks, as a broader attack class relevant to all three regulatory risks.
Failing to mitigate them under realistic assumptions can compromise protection against all three risks, as an adversary could fully reconstruct and observe all attributes of an individual in the training data.
If an adversary can reconstruct even a handful of targets with high confidence, this constitutes a serious privacy breach, effectively rendering the model non-private~\cite{carlini2022membership,carlini2023extracting}.
However, notwithstanding the importance of defending against reconstruction attacks, this alone is not sufficient to address all regulatory risks, as other attacks (like those mentioned above) can still expose sensitive information even without full record reconstruction.

\section{Does Synthetic Data Without Privacy Mechanisms Provide Sufficient Anonymization?}
\label{sec:anon}
In this section, we examine whether synthetic data can be considered sufficiently anonymized from a data protection perspective.
Overall, we discuss why, without additional privacy defenses, it patently does not (as per the main privacy attacks in Table~\ref{tab:reg}).

As mentioned, sampling synthetic datasets through generative models %
may appear to address privacy concerns, e.g., by breaking the direct one-to-one mapping from real to synthetic records, potentially making linkability more difficult~\cite{cohen2022attacks,giomi2022unified}.
Additionally, the inherent stochastic nature of generative models ensures that each synthetic record is produced with some degree of randomness, leading to uncertainty and variability across generated datasets.
This randomness can help obscure unique or rare data points, reducing the likelihood of singling out individuals.
In fact, the EC's Joint Research Centre has argued that synthetic data makes re-identification ``virtually impossible''~\cite{girolamo2024synthetic}.\footnote{``As opposed to traditional anonymization techniques, the synthesis process makes any potential attack virtually impossible. This is because instead of keeping an anonymized version of the original data, the synthesis strips it to a limited set of parameters which is then used to build up a completely new dataset with similar statistical properties. This makes any identification of individual original observations impossible, because they are simply no longer there''~\cite{girolamo2024synthetic}.}

However, this perspective overlooks critical vulnerabilities.
As discussed in Section~\ref{subsec:attacks}, generative models may overfit to specific training records, potentially recreating them exactly or approximately.
Under the motivated intruder test, %
adversaries with simple black-box access to the model can exploit overfitting and extract sensitive information about the training data.
For instance, adversaries can infer with high certainty whether certain target records were part of the training data via MIAs~\cite{van2023membership,chen2020gan,hayes2019logan,stadler2022synthetic}, thereby severely increasing the linkability risk.
Furthermore, they can accurately predict unknown attributes of arbitrary records due to individual-level information leakage and overly accurate aggregate statistics (in some cases with up to 95\% probability)~\cite{annamalai2024linear}, exposing the training data to inference risks.
Adversaries could also single out entire training records by reconstructing them by repeatedly sampling the trained model~\cite{carlini2023extracting,carlini2021extracting,dick2023confidence}, thus failing another key risk.

\descr{\em{Take-away.}}
Without additional privacy protections, synthetic data techniques alone fail to address the three key regulatory risks adequately; consequently, models and data should thus not be considered sufficiently anonymized and, thus, not anonymous data~\cite{bellovin2019privacy,lopez2022on}.

\section{Does Synthetic Data With Privacy Mechanisms (DP vs. SBPMs) Provide Sufficient Anonymization?}
\label{sec:anondp}
We now analyze and compare the two most common mechanisms for synthetic data, DP and SBPMs, first from a privacy perspective and then from a regulatory perspective, assessing whether they can satisfy anonymization requirements.
We report the main privacy attacks and the assumptions we use for our analysis in Table~\ref{tab:reg}.

For completeness, we also compare DP and SBPMs according to six non-privacy-related criteria that are nevertheless important in practice and present edge-case examples illustrating contrasting behaviors to build intuition.

Overall, we discuss in detail how combining synthetic data with DP or with SBPMs leads to very different levels of protection.
DP provides formal, worst-case guarantees and bounds leakage even against powerful future attacks, though typically at the cost of reduced utility, fairness concerns, and higher implementation/interpretation complexity.
SBPMs are simple, intuitive, and fast while preserving utility, but they lack formal guarantees, rely on average-case statistics, and can leak sensitive records while still passing their tests.

\subsection{DP vs.~SBPMs: Privacy Perspective}
\label{subsec:priv}
We introduce eight privacy-related criteria and analyze DP and SBPMs vis-\`a-vis these criteria.
A summary is shown in Table~\ref{tab:priv}.

\begin{table*}[t!]
	\centering
  \begin{tabular}{lcc}
    \toprule
		\bf Privacy Criterion 			& \bf DP						 				& \bf SBPMs									\\
    \midrule
    Adversarial Model  					& well-defined      				& undefined		  						\\
		Privacy Guarantees    		 	& general  									& empirical		  						\\
		Privacy Analysis					 	& worst-case				 				& average-case							\\
		Privacy Risk    					 	& overestimation     				& underestimation 					\\
		Plausible Deniability			 	& yes								 				& no												\\
		Privacy Subject			  		 	& generative model 					& synthetic data						\\
		Privacy Leakage							& continuous								& binary										\\
		Privacy Expenditure  				& training time             & generation time           \\
    \bottomrule
	\end{tabular}
	\caption{Comparison between synthetic data with privacy guaranteed by DP and SBPMs across eight privacy-related criteria.} %
	\label{tab:priv}
\end{table*}

\descr{1.~Defined Adversarial Models.} %
We begin by %
examining the strategic adversaries DP and SBPMs consider~\cite{anderson2020security}.
As explained in Section~\ref{subsec:DP}, DP assumes well-defined adversaries; depending on the specific DP definition~\cite{desfontaines2020sok}, they are allowed to add/remove/modify a single record in the training data but are still unable to infer private information about that record (or others), up to a certain probability.
These protections remain effective even in the presence of a very powerful adversary that, along with white-box access to the model, can even observe and intervene in intermediate training steps~\cite{annamalai2024what,jagielski2020auditing,nasr2021adversary} and possess other side information.
However, this comes with an inherent tradeoff -- stronger privacy usually requires more noise/randomness, which may reduce utility (see Section~\ref{subsec:add}).

By contrast, SBPMs do not typically define a formal adversarial model.
Rather, they focus solely on the results of statistical tests comparing the train-test and train-synthetic distances; if the tests pass, the synthetic data is assumed to be private.
However, this approach implicitly allows for maliciously crafted records to be included in the synthetic data (as long as the tests pass), leaving SBPMs vulnerable to various attacks~\cite{ganev2025inadequacy}, as we show in Section~\ref{subsec:ex}.

\descr{2.~Privacy Guarantees.}
DP defines theoretical guarantees regarding the information leakage associated with any single record in the training data.
These guarantees are future-proof, as they hold against both current and future privacy attacks~\cite{dwork2014algorithmic,wood2020differential}.
Interpreting their strength in practice remains challenging, however, as selecting an appropriate privacy budget is still an open problem.
Still, several measurement studies empirically show that applying DP does reduce the privacy leakage from membership, attribute, and reconstruction attacks~\cite{annamalai2024linear,giomi2022unified,houssiau2022tapas,stadler2022synthetic}. %
By contrast, SBPMs %
only focus on the privacy risks they measure. %
As discussed in Section~\ref{subsec:emp}, SBPMs do not perform well empirically, even in defending against the risks they are designed to measure~\cite{annamalai2024linear,ganev2025inadequacy,houssiau2022tapas,yao2025dcr}.

\descr{3.~Privacy Analysis.}
DP reasons about protections for any target and any (defined) neighboring datasets, under strong adversarial assumptions.
Thus, it provides worst-case analyses, providing privacy for every individual record in the training data, regardless of how different it may be from the others.

On the other hand, SBPMs focus on average-case statistics, e.g., comparing metrics like the mean and the 5th percentile of train-test and train-synthetic distances.
As a result, SBPMs prioritize privacy protection for the ``average-looking'' record, while potentially overlooking the privacy risks for outliers.
Overall, SBPMs fall into the {\em ``generalization implies privacy''} fallacy~\cite{del2023bounding}, where generalization is an average-case issue but privacy is a worst-case.
In fact, even if a machine learning model passes all tests and generalizes, it can still memorize data~\cite{song2017machine}.

\descr{4.~Privacy Risk.}
DP's level of privacy protection is controlled by the privacy budget parameter, which serves as a theoretical upper bound (or maximum amount) on information leakage.
Approaches to empirically validate these guarantees typically involve running MIAs and estimating privacy leakage based on the adversary's accuracy~\cite{annamalai2024what,houssiau2022tapas}.
However, to achieve {\em ``tight''} empirical estimates (i.e., estimates close to the theoretical bounds), adversaries often need to be given very strong assumptions, such as white-box access and training the model with two or three records alongside a highly distinct target record~\cite{annamalai2024what}.
In practice, these assumptions may be too strong for real-world scenarios, leading to the argument that DP may overestimate realistic privacy risks.

SBPMs operate with purely empirical estimates for the specific privacy risks they aim to quantify, meaning that the actual risk is equal to or higher than what is measured.
As highlighted by researchers~\cite{desfontaines2024empirical,ganev2025inadequacy,yao2025dcr}, the actual privacy risk is often significantly higher and cannot be bounded by SBPMs.
While their privacy risk scores are arguably easy to compute and interpret, they tend to underestimate the real risk.

\descr{5.~Plausible Deniability.}
DP provides plausible deniability through contrastive analysis, defining the output to be (nearly) indistinguishable whether or not a specific record was used in the model training.
Therefore, even if a generated record is an exact or approximate copy of a training record, %
DP protection still holds.
The individual can plausibly deny that their data was used, as DP %
makes it statistically possible that the record was generated by chance.
This is a particularly desirable property, especially in highly sensitive settings where severe consequences could arise, e.g., in the case of the United Nation's International Organization for Migration (IOM) %
releasing a DP synthetic dataset on victim-perpetrator relations to combat human trafficking~\cite{microsoft2020iom}.

By contrast, SBPMs do not consider comparisons with or without an individual's data.
Without adding noise or randomness, this approach eliminates plausible deniability, allowing adversaries to be certain when leaking information~\cite{ganev2025inadequacy}.
Additionally, exact replication of training records is directly reflected in one of the three privacy metrics, namely Identical Match Share (IMS) (see Section~\ref{subsec:sbpms}).

\descr{6.~Privacy Subject.}
Differentially private synthetic data generation algorithms satisfy DP during the training phase of the model.
Thus, the primary privacy subject is the trained generative model and its parameters/weights.
Even with white-box access to the model, an adversary cannot extract more information than allowed by DP's guarantees.
In other words, privacy becomes a property of the computation, not of any specific output data, as argued by Nissim et al.~\cite{nissim2018bridging}.\footnote{``A benefit of viewing privacy as a property of computation is that computations are formal mathematical objects and as such are amenable to rigorous mathematical arguments [...] In turn, these arguments can assure actors who release data that they are acting in accordance with the law''~\cite{nissim2018bridging}.}
Furthermore, thanks to DP's post-processing property, an arbitrary number of synthetic datasets can be generated from the trained model without incurring additional privacy risks.

On the other hand, SBPMs treat privacy as a property of the synthetic data, rather than the generative model or process.
This approach requires running the privacy tests each time synthetic data is generated, introducing potential additional privacy leakage with every new dataset---we discuss the consequences in point 8.~below.

\descr{7.~Information Leakage.}
DP treats leakage as a continuous value controlled by the privacy budget parameter.
This allows for a flexible balance between privacy and utility, aligning with the ``spectrum of identifiability'' outlined by the ICO~\cite{ico2021how}.
Whereas SBPMs handle leakage in a binary manner: the synthetic data is either deemed private if it passes the three privacy tests or considered non-private if it fails any.
This binary approach can be undesirable, as it lacks transparency regarding what factors influence the tests' outcomes.

\descr{8.~Privacy Expenditure.} %
As previously discussed, DP generative models allow the generation of an arbitrary number of synthetic datasets from a fixed DP-trained model without additional privacy accounting, as subsequent sampling is covered by DP's post-processing property.
This includes downstream models trained solely on DP synthetic data, while retraining/fine-tuning on sensitive data incurs additional privacy loss that must be tracked through composition.
This is not the case for SBPMs, as the metrics must be recalculated for each generated synthetic dataset.
Since SBPMs treat privacy leakage in a binary manner, they assume that releasing one synthetic dataset is as safe as releasing multiple datasets (as long as the tests pass), even though the train-synthetic distances need to be computed for each release.

However, as per the {\em ``Fundamental Law of Information Reconstruction''}~\cite{dwork2014algorithmic}, too many accurate answers can severely compromise privacy, a risk SBPMs fail to account for.
Indeed, this has been exploited to run successful privacy attacks against SBPMs~\cite{ganev2025inadequacy}.

\subsection{DP vs~SBPMs: Regulatory Perspective}
\label{subsec:pm}
We now discuss whether combining synthetic data generation with DP or SBPMs can be used to meet the regulatory anonymization requirements.

\descr{Synthetic Data with DP.}
First, DP was specifically designed with plausible deniability protections (via randomness and noise addition), which ensures that trained models/synthetic datasets are statistically indistinguishable when the input dataset differs by a single record.
These probabilistic guarantees are consistent with the risk-based anonymization standard discussed in Section~\ref{subsec:reg}, which does not require identifiability risk to be eliminated entirely.
DP's indistinguishability guarantees naturally minimize the impact of differencing attacks, as adversaries cannot reliably determine if an individual's data is included, thus reducing singling out concerns.
However, as mentioned in Section~\ref{subsec:emp}, protecting against differencing attacks does not necessarily address all singling-out risks.
To this end, Cohen and Nissim~\cite{cohen2020towards} provide a robust formalization of DP's protections against predicate singling out, an attack that aligns with GDPR's singling out concept.
Furthermore, empirical studies~\cite{giomi2022unified} demonstrate the effectiveness of DP mechanisms in mitigating singling out risks when only observing the synthetic data.

Second, DP guarantees reduce linkability risks by limiting the influence of individual training records on the model.
This has been empirically demonstrated in several studies, where researchers have modeled one manifestation of linkability risk using MIAs~\cite{van2023membership,giomi2022unified,stadler2022synthetic,kumar2020mlprivacy} in black-box settings.
The ICO also suggests using DP with synthetic data to protect the privacy of outliers or other vulnerable individuals~\cite{ico2023privacy}, a protection empirically confirmed by Stadler et al.~\cite{stadler2022synthetic}.
Moreover, DP protections mitigate the vulnerability to MIAs even in powerful white-box scenarios and worst-case conditions~\cite{annamalai2024what}, which are arguably unrealistic assumptions for most real-world adversaries~\cite{cummings2024attaxonomy}.

Next, DP mechanisms can reduce inference risks, limiting additional leakage from individual-level information -- this was shown both in black-box settings~\cite{giomi2022unified,stadler2022synthetic} and stricter no-box settings~\cite{annamalai2024linear}, where the adversary only has access to the synthetic dataset.
Moreover, they can mitigate reconstruction attacks, which recover: 1) individual records from accurate statistics or aggregate queries (such as 2-way and 3-way marginals) on large populations~\cite{dick2023confidence}, and 2) an unknown target record from a trained classifier when exact knowledge of all other records is available~\cite{balle2022reconstructing}.

Finally, recent theoretical work directly connects DP guarantees to attack-based privacy risks, including bounds on reconstruction attacks~\cite{balle2022reconstructing,hayes2023bounding} and unified frameworks covering re-identification, attribute inference, and reconstruction~\cite{kulynych2025unifying,guerra2026understanding}, providing theoretical support for interpreting DP guarantees through concrete privacy attacks.

\descr{\em{Take-away.}}
Although implementing DP algorithms remains technically challenging and their privacy and utility effectiveness are highly context-dependent, when properly applied, DP synthetic data generation can reduce all three regulatory identifiability risks to sufficiently low levels so that models and synthetic datasets can be considered anonymous, with both theoretical and empirical support.

\descr{Synthetic Data with SBPMs.}
By contrast, SBPMs are neither an adequate mechanism for measuring the privacy of synthetic data nor for mitigating the three regulatory risks.
First, they do not offer plausible deniability protections, one of many crucial shortcomings discussed in Section~\ref{subsec:priv}, and they are susceptible to differencing attacks, which increase singling-out risks.
Even for the risk they were primarily designed to address -- i.e., linkability, which Distance to Closest Record (DCR) loosely attempts to emulate -- DCR significantly underestimates membership inference risks in black-box settings~\cite{annamalai2024what,ganev2026smote,houssiau2022tapas,yao2025dcr} compared to popular MIAs.

Moreover, an adversary with minimal black-box access (i.e., to a single trained model and the metrics) can launch successful MIAs and AIAs with just a few API calls to the metrics~\cite{ganev2025inadequacy}.
Most concerningly, SBPMs are vulnerable to simple reconstruction attacks~\cite{ganev2025inadequacy}, which invalidates their overall privacy protections and arguably makes them unsuitable to meet regulatory requirements.
The FCA also notes that SBPMs are ineffective at protecting privacy concerns~\cite{fca2024using}.\footnote{``[...] we note, that distance-based privacy metrics alone are known to be ineffective at protecting all privacy concerns, for synthetic data. As highlighted by recent research, membership inference attacks can be valuable approach to evaluate privacy and identify where data may not be adequately protected against from a privacy perspective''~\cite{fca2024using}.}

\begin{table*}[t!]
	\centering
  \begin{tabular}{lcc}
    \toprule
		\bf Consideration 					 & \bf DP						 				& \bf SBPMs						\\
    \midrule
    Utility  									 	 & reduced      					  & not affected		  	\\
		Fairness    		 						 & reduced/disparate  			& not affected		  	\\
		Consistency			 						 & not always								& no									\\
		Interpretation    					 & challenging     					& seems intuitive	    \\
		Computational Performance		 & slow											& fast								\\
		Implementation and Adoption	 & difficult 								& easy								\\
    \bottomrule
	\end{tabular}
	\caption{Comparison between synthetic data with privacy guaranteed by DP and SBPMs across six non-privacy-related criteria.}
	\label{tab:add}
\end{table*}

\descr{\em{Take-away.}}
SBPMs may be intuitive, easy to implement, and preserve utility, but they greatly underestimate the privacy risks of synthetic data generation and fail to detect leakage even in basic scenarios (as shown in Section~\ref{subsec:ex}).
Consequently, they do not provide a reliable mechanism for reducing regulatory risks or ensuring anonymization.

\subsection{DP vs~SBPMs: Non-Privacy Considerations}
\label{subsec:add}
We now discuss additional practical considerations beyond privacy for the two mechanisms.
We also provide a concise comparative summary in Table~\ref{tab:add}.

\descr{Utility.}
With DP, the privacy level can be controlled via the desired privacy budget (i.e., the $\epsilon$ parameter).
Naturally, the higher the privacy, the more noise is added to the process, which yields an inherent utility-privacy tradeoff.
This tradeoff is not easily predictable and depends on various factors, e.g., the size and shape of the training data, the complexity of the downstream task, the choice of the generative model, the specific DP mechanism, etc.~\cite{ganev2024graphical,jordon2022synthetic,tao2022benchmarking}.
There is no universally agreed-upon privacy budget (nor an attack-success/advantage threshold) among researchers, policymakers, and regulators that would guarantee privacy concerns are fully addressed; however, in many real-world deployments, the privacy budget tends to fall between 1 and 20~\cite{desfontaines2021list}.
In some cases, where privacy risk tolerance needs to be very conservative, e.g., when sharing highly sensitive genomic data, current solutions may still fall short of delivering acceptable performance~\cite{oprisanu2021utility}.

SBPMs, on  the other hand, focus solely on the synthetic data when measuring privacy at generation.
As a result, they do not necessarily require lowering utility in order to pass the privacy tests.
However, if a test initially fails, it is difficult to determine whether reduced utility would help, and it is unclear which specific aspects or hyperparameters of the model should be adjusted, if possible. %

\descr{Fairness.}
Utility degradation in DP synthetic data is usually non-uniform, i.e., outliers and underrepresented subgroups tend to suffer from bigger drops compared to ``average-looking'' records and majority subgroups~\cite{bagdasaryan2019differential,cheng2021can,ganev2022robin,kulynych2023arbitrary}.
This is expected and inherently hard to mitigate, as protecting outliers requires injecting comparatively more noise.
Evidence of this issue has already emerged in the US Census, where DP disproportionately distorts counts for small and minority communities, raising concerns about misrepresentation and unfair funding~\cite{kenny2021use,kuppam2020fair}.
Reducing this uneven utility degradation remains an area of active research~\cite{rosenblatt2024simple,vero2024cuts}.
By contrast, in principle, SBPMs do not disproportionately affect the utility of outliers. %

\descr{Consistency.}
Even though most DP generative models are consistent, not all are; that is, they do not necessarily become more accurate with additional training data~\cite{ganev2024graphical}.
Additionally, different models are better suited for different data characteristics/downstream tasks.
Therefore, practitioners should exercise caution when training synthetic data models.

SBPMs exhibit inconsistency in a different way: their privacy scores depend heavily on randomness in both the train/test split and the generation process.
Specifically, even with fixed training and test datasets, sampling from the same trained model can yield highly variable statistical test results.
This inconsistency also occurs with fixed synthetic data and different splits between training and test data~\cite{ganev2025inadequacy}.

\descr{Interpretation.}
While DP offers theoretical guarantees, its protections can be difficult to interpret and communicate to non-experts, especially when practitioners or policymakers must decide whether a given privacy budget $\epsilon$ is ``safe enough'' for their application~\cite{nanayakkara2023chances}.
Understanding DP also requires reasoning about neighboring datasets, privacy unit, and composition across model retraining/fine-tuning, DP variant/mechanism, worst-case adversaries, etc. -- concepts that are often misunderstood outside the DP research community~\cite{cummings2021need,houssiau2022on,nanayakkara2023chances,nasr2021adversary,song2024inherently,tang2017privacy}.
Recent work has partly addressed this interpretability challenge by translating DP guarantees into concrete bounds on membership, attribute inference, and reconstruction risks~\cite{balle2022reconstructing,hayes2023bounding,kulynych2025unifying,guerra2026understanding}.
Nonetheless, selecting and communicating an appropriate level of privacy protection remains context-dependent, and DP proponents argue that such complexity is inherent to any meaningful privacy protection and should not hinder the adoption of well-founded mechanisms~\cite{cyffers2025setting}.

On the other hand, SBPMs appear more intuitive and easier to interpret at first, as passing their statistical tests can be mistakenly understood as evidence that synthetic data is private.
However, such results do not necessarily guarantee privacy: failing to reject the alternative hypothesis does not imply acceptance of the null.
In practice, the generative process or synthetic data may still be vulnerable to untested privacy attacks (as argued and demonstrated by~\cite{desfontaines2024empirical,ganev2025inadequacy,yao2025dcr}).

\descr{Computational Performance.}
DP-SGD implementations~\cite{abadi2016deep} -- the most widely used framework for training or fine-tuning deep learning models with DP -- can be significantly slower than standard (non-private) training~\cite{subramani2021enabling,yousefpour2021opacus}.
The main reason is that DP-SGD requires computing per-example gradients, which is more computationally intensive than the usual aggregated gradient computation.
SBPMs are much faster to compute since they operate on the generated dataset rather than modifying the training process.

\begin{figure*}[t!]
\centering
\begin{subfigure}[b]{0.99\linewidth}
	 \centering
	 \includegraphics[width=0.5\textwidth]{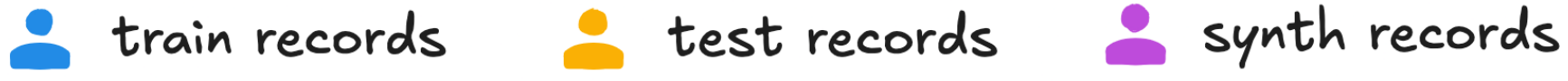}\vspace{0.2cm}
\end{subfigure}
\begin{subfigure}[b]{0.25\linewidth}
	 \centering
	 \includegraphics[width=0.9\textwidth]{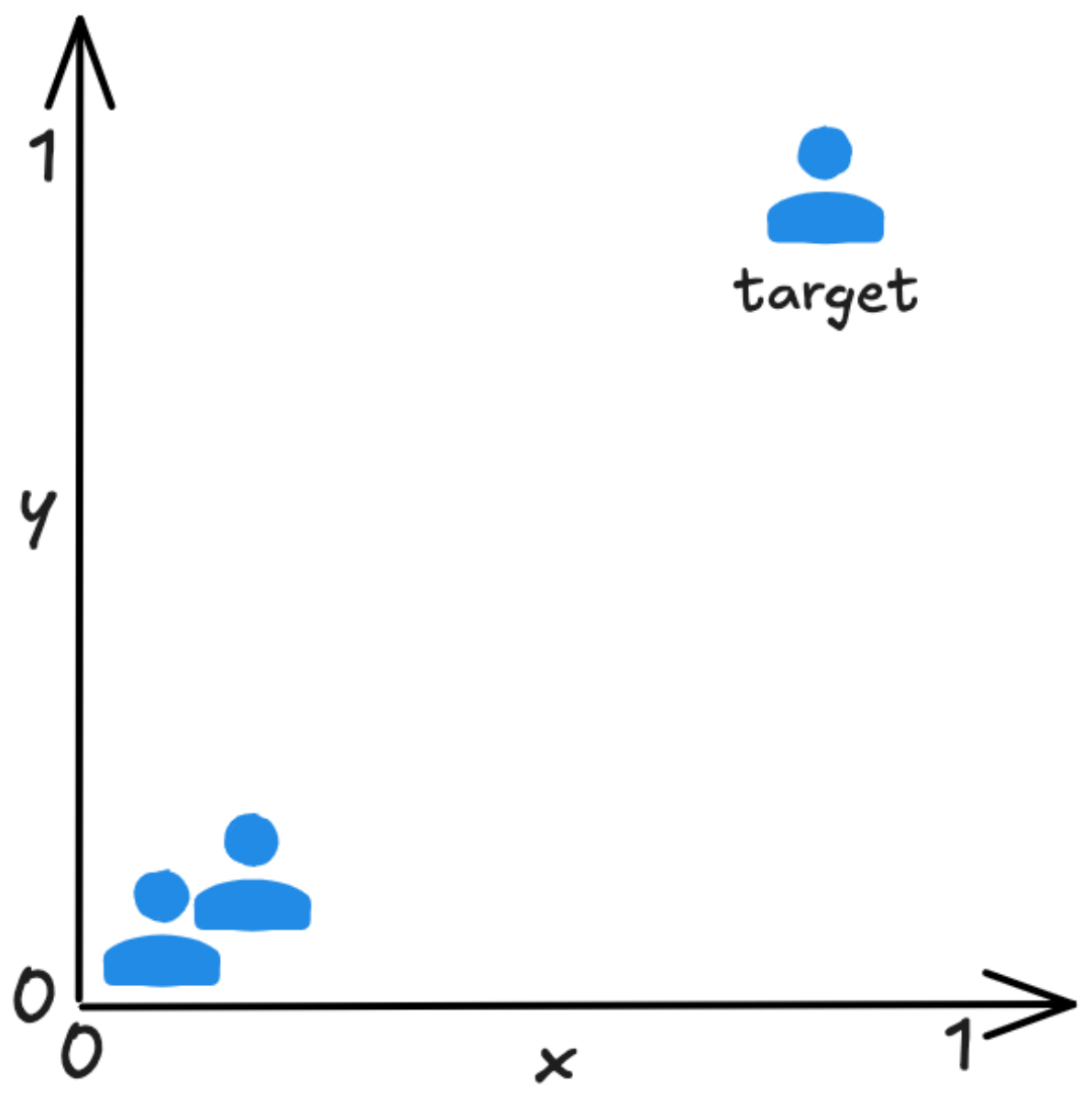}
	 \caption{DP guarantees applying for worst-case target record and worst-case neighboring datasets}
	 \label{fig:ex_df}
\end{subfigure}
\hfill
\begin{subfigure}[b]{0.35\linewidth}
	 \centering
	 \includegraphics[width=0.9\textwidth]{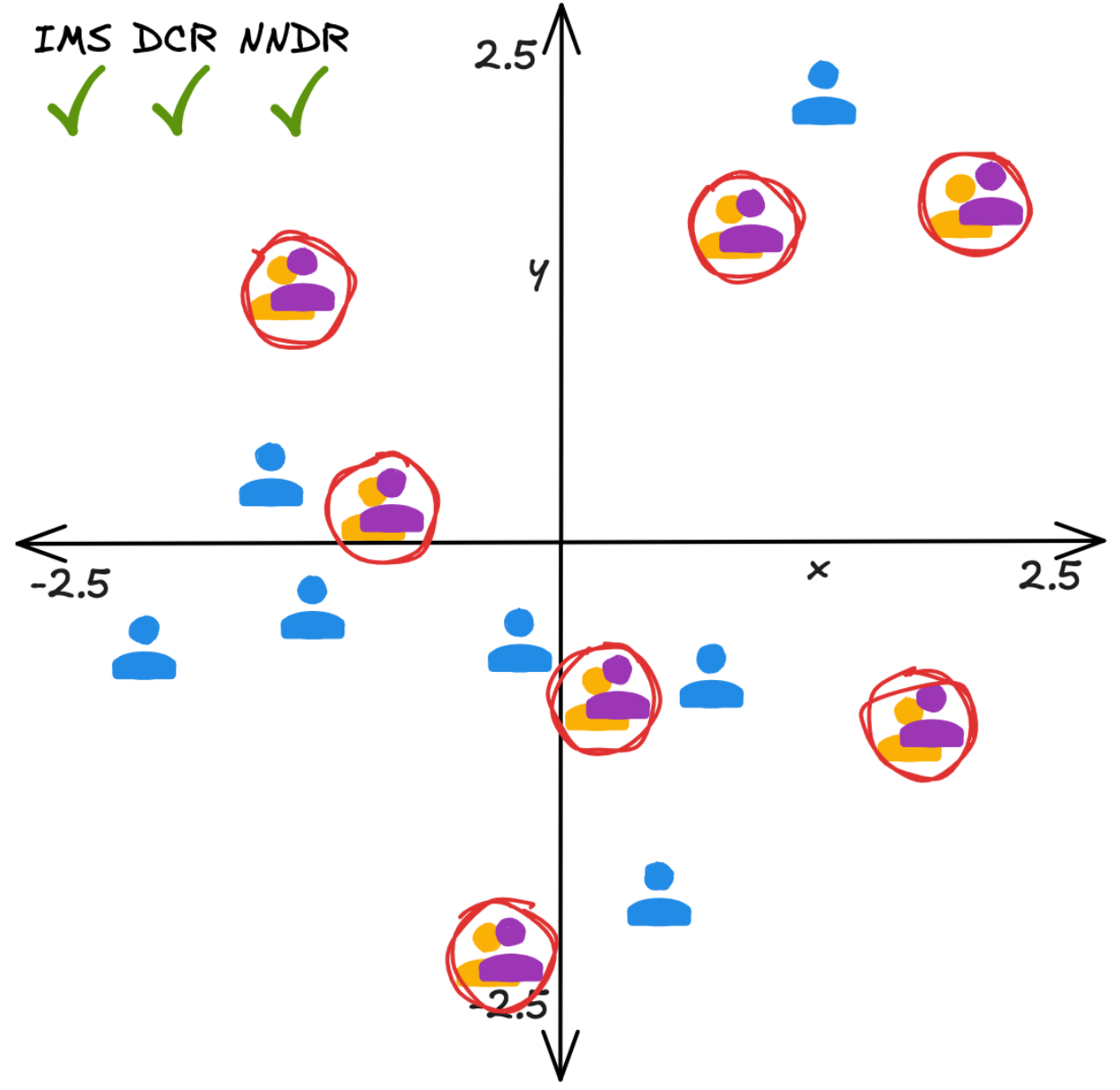}
	 \caption{SBPMs leaking all test records (in red circles) even when all statistical tests pass\\}
	 \label{fig:ex_sbpm_test}
\end{subfigure}
\hfill
\begin{subfigure}[b]{0.35\linewidth}
	 \centering
	 \includegraphics[width=0.9\textwidth]{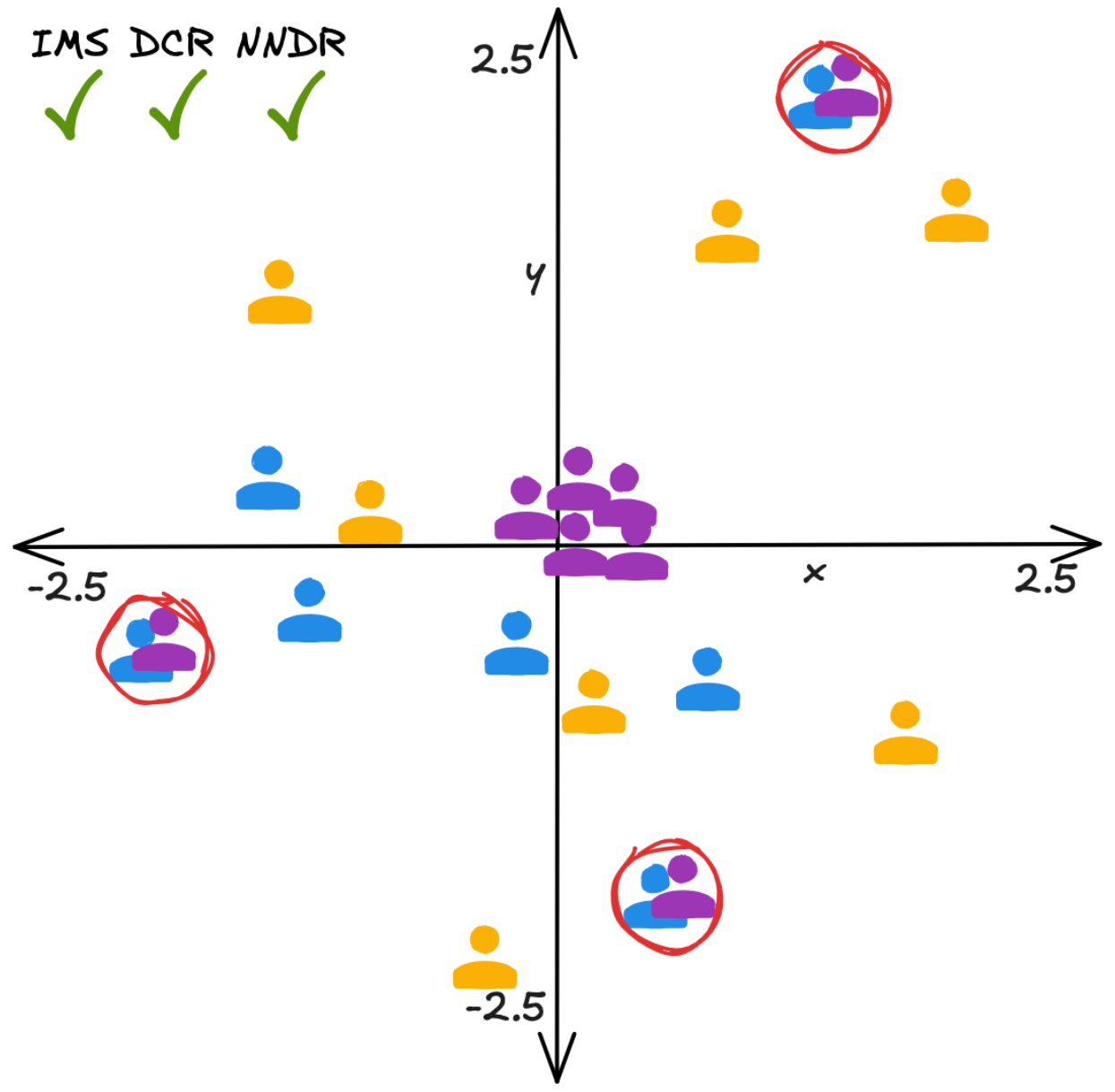}
	 \caption{SBPMs leaking all training outliers (in red circles) even when all statistical tests pass\\}
	 \label{fig:ex_sbpm_train}
\end{subfigure}
\caption{Base-case examples of synthetic data with privacy guaranteed by DP and SBPMs.}
\label{fig:ex}
\end{figure*}

\descr{Implementation and Adoption.}
Implementing and deploying DP generative models remains challenging and requires specialized technical expertise.
As already discussed, and as corroborated by recent NIST DP evaluation guidelines~\cite{near2025guidelines}, incorporating DP requires several non-trivial decisions, like selecting an appropriate privacy budget, defining the privacy unit (e.g., user/document/example-level) and neighboring datasets, etc.
In practice, achieving usable utility may require relatively large privacy budgets, which risks negating meaningful protection.
In limited data domains (e.g., tables with a handful of columns and small category sets), even moderate privacy budgets can cause substantial leakage~\cite{drechsler2023differential}.
Moreover, as with other security primitives, practitioners are discouraged from designing custom DP mechanisms and are instead advised to rely on vetted, publicly available implementations.
Yet, even well-scrutinized DP libraries can contain subtle bugs~\cite{annamalai2024what,desfontaines2025dont,ganev2025elusive,cebere2026privacy,haney2022precision,lokna2023group} that are difficult to detect and may silently compromise privacy.
Outside of the US Census, only a few large technology companies have published detailed, real-world DP deployments~\cite{desfontaines2021list,opendp2025differential}, highlighting both the difficulty and expertise needed for their adoption.

Unlike DP, SBPMs require no privacy parameters, no explicit adversarial model, and no utility/fairness tradeoffs.
They are typically computationally lightweight and produce results that seem straightforward to interpret -- factors that make them popular among synthetic data vendors.
However, even if a single metric is easy to implement, selecting from the dozens of SBPM variants~\cite{boudewijn2023privacy} is poorly understood, and the implications of each choice remain uncertain.
Despite their prevalence among synthetic data vendors, SBPMs are not adopted by large technology companies.

\subsection{DP vs.~SBPMs: Base-Case Examples}
\label{subsec:ex}
To provide intuitions as to how the privacy mechanisms protect (or not) sensitive data in simple scenarios and inspired by~\cite{annamalai2024what,ganev2025inadequacy,lokna2023group}, we construct three base cases, visualized in Figure~\ref{fig:ex}.
While these scenarios are intentionally extreme and seem unrealistic, they arguably serve as simple examples for testing or auditing the privacy properties of mechanisms.
Somewhat similar to unit tests in software development, these examples assess privacy protections under worst-case conditions, which could materialize under especially powerful threat models.

\descr{DP: Worst-case.}
As discussed earlier, DP guarantees hold in worst-case scenarios.
This includes cases where the training data consists of only two records, and the target record is an outlier, as shown in Figure~\ref{fig:ex_df}.
In fact, similar scenarios have been used by~\cite{annamalai2024what,lokna2023group} to tightly estimate the privacy budget of DP algorithms empirically.
However, any DP generative model trained on such limited data would likely produce results with little utility.

\descr{SBPMs: Leaking all Test Records.}
As originally reported by~\cite{ganev2025inadequacy}, Figure~\ref{fig:ex_sbpm_test} shows how severe privacy leakage can still occur even if all SBPMs statistical tests pass.
Here, the training and test data both contain ten records randomly sampled from a 2d standard normal distribution, while the synthetic data is an exact replica of the test data.
Obviously, publishing half of the personal records in this manner can be considered neither private nor regulatory-compliant.

\descr{SBPMs: Leaking all Training Outliers.}
Another scenario, again inspired %
by~\cite{ganev2025inadequacy}, where privacy leakage can occur despite all SBPMs passing, is illustrated in Figure~\ref{fig:ex_sbpm_train}.
Using the same train/test data as in the previous example, we construct synthetic data that includes three train records considered outliers (located farther away from the origin) with small perturbations, along with 70 copies of the value (0, 0).
Publishing this synthetic data would leave the individuals represented by the outlier records unconvinced that their privacy is preserved~\cite{ons2018privacy}, particularly in the absence of plausible deniability (which is ruled out by SBPMs).
This concern is especially relevant given that the ICO has emphasized the importance of protecting outliers in privacy assessments~\cite{ico2022privacy}.

\section{Related Work}

\descr{Synthetic Data.}
Limited research has focused on the intersection between synthetic data and regulation.
Bellovin et al.~\cite{bellovin2019privacy} discuss limitations of synthetic data without additional privacy mechanisms, suggesting that DP can help address these concerns.
However, they do not explore privacy attacks and their analysis is limited to narrow regulations like HIPAA and FERPA rather than the GDPR.

In the context of the GDPR, legal experts~\cite{lopez2022on} have argued that synthetic data could be considered personal, pseudonymous, or anonymous depending on its level of identifiability, while others~\cite{batista2023synthetic} assert it should be viewed as anonymous.
However, they give little consideration to vulnerabilities from privacy attacks and overlook both modern synthetic data workflows and available privacy mechanisms.
Gal and Lynskey~\cite{gal2023bridging} explore the potential impact of synthetic data on data flows, governance, and privacy laws.
They discuss the governance challenges prompted by synthetic data and challenge the claim that synthetic data does not constitute personal data, raising fundamental questions as to whether data that is not linked to an individual in the training data should still be considered personal data and how inferences based on collected data should be treated.
Similarly, Whitney and Norman~\cite{whitney2024real} examine how synthetic data can be used to circumvent consent requirements and obscure data lineage, posing regulatory challenges for enforcement agencies like the US Federal Trade Commission, which relies on consent to request model deletion.
Prior work has also highlighted how privacy-enhancing technologies like synthetic data can be misused for ``privacy washing,'' i.e., creating the appearance of compliance while avoiding accountability and leaving underlying risks unaddressed~\cite{de2025pets,yew2025anti}.

\descr{Privacy Metrics.}
Privacy metrics have also been proposed that map to the three regulatory risks within the context of GDPR~\cite{giomi2022unified}.
However, these metrics rely on distances between specific train/test/synthetic samples rather than the generative model itself, which has been shown to greatly underestimate actual risk~\cite{annamalai2024what,houssiau2022tapas,yao2025dcr}.
As a consequence, Giomi et al.~\cite{giomi2022unified} suggest that the risks of using synthetic data without privacy mechanisms are generally very low (especially linkability), and that DP provides only marginal improvements.
This aligns with recent work showing that the linkability metric barely detects any leakage even in clearly non-private generative models~\cite{ganev2026smote}.
In contrast, we highlight that synthetic data techniques remain vulnerable to a range of privacy attacks targeting the generative model.
We further analyze the effectiveness of these attacks under varying levels of assumption strictness and assess their impact on commonly used privacy mechanisms.

Similarly, Desfontaines~\cite{desfontaines2024empirical} argues that privacy metrics often underestimate the very risks they aim to measure, highlighting the need for more rigorous privacy evaluation standards for both users and regulators of synthetic data technologies.

\section{Conclusion}
In this paper, we connected personal data regulations in the EU and UK to synthetic data generation techniques through the lens of privacy attacks.
We interpreted the regulatory identifiability risks that must be mitigated in the context of generative machine learning and synthetic data, arguing that these risks should be assessed using a model-centric perspective and systematically evaluated via state-of-the-art privacy attacks.
Our analysis of current deployments shows that combining synthetic data generation with DP guarantees can effectively mitigate these risks, whereas relying on SBPMs provides insufficient protection.

\descr{Limitations and Future Work.}
While relying on privacy attacks and DP offers a more robust approach to privacy, implementing and configuring these techniques can be challenging for non-experts and often introduce additional computational overhead.
Moreover, the extent to which DP mitigates regulatory identifiability risks is highly context-dependent, influenced by various factors, making it difficult to provide universal guidelines.
As part of future work, we are confident that developing a broader range of efficient privacy attacks could provide deeper insights into the behavior and vulnerabilities of synthetic data approaches.
This could also help formalize the relationship between regulations and privacy attacks across different assumptions and settings.

\descr{Ethical Considerations Statement.}
We believe our analysis has important implications for the broader community.
Our findings encourage machine learning researchers, practitioners, and policymakers to adopt model-centric risk assessments that evaluate privacy risks at the level of the generative model itself and highlight the need for closer cross-disciplinary collaboration.
Our work also supports the continued adoption of end-to-end DP pipelines, which offer strong, principled protections against motivated adversaries and help align synthetic data products with anonymization requirements.
At the same time, clients and downstream users of these technologies must invest in education, critically reassess the privacy risks of ongoing projects, and demand more rigorous evaluation standards.

\descr{Acknowledgements.}
We are grateful to Orla Lynskey and Paul Comerford (acting in a personal capacity) for their valuable feedback and suggestions, which helped us significantly improve our paper.
\bibliographystyle{ACM-Reference-Format}
\input{no_comments.bbl}

%

%
%
%

\end{document}